\documentclass[instruments,article,accept,moreauthors,pdftex]{Definitions/mdpi} 
\graphicspath{{/}}

\firstpage{1} 
\pubvolume{xx}
\issuenum{1}
\articlenumber{5}
\pubyear{2020}
\copyrightyear{2020}
\history{Received: 22 June 2020; Accepted: 10 July 2020; Published: 14 July 2020}
\updates{yes} 
\usepackage{upgreek}
\usepackage{subfigure}
\makeatletter
\renewcommand{\@thesubfigure}{\normalsize(\textbf{\alph{subfigure}})}
\makeatother

\usepackage{color}
\usepackage{subfigure}
\usepackage[para,flushleft]{threeparttable}

\Title{Infrared Wavefront Sensing for Adaptive Optics Assisted Galactic Center Observations with the VLT Interferometer and GRAVITY: Operation and Results}

\Author{{Stefan Hippler} 
 $^{1,}$*\orcidA{}, Wolfgang Brandner $^{1}$, Silvia Scheithauer $^{1}$, Martin Kulas $^{1}$, 
Johana~Panduro~$^{1}$, Peter Bizenberger $^{1}$, 
Henry Bonnet $^{2}$, Casey Deen $^{3}$, Fran\c{c}oise~Delplancke-Str\"{o}bele $^{2}$, Frank~Eisenhauer $^{3}$,
Gert Finger $^{2}$, Zoltan Hubert $^{4}$, Johann~Kolb~$^{2}$, Eric~M\"{u}ller ~$^{2}$, Laurent~Pallanca $^{2}$,
Julien Woillez $^{2}$, G\'{e}rard Zins $^{2}$ and {GRAVITY Collaboration} $^{3}$
}

\address{%
$^{1}$ \quad Max-Planck-Institut f\"{u}r Astronomie, 69117 Heidelberg, Germany; brandner@mpia.de (W.B.); scheithauer@mpia.de (S.S.); kulas@mpia.de (M.K.); panduro@mpia.de (J.P.); biz@mpia.de (P.B.) \\
$^{2}$ \quad European Southern Observatory Headquarters, 85748 Garching, Germany; hbonnet@eso.org (H.B.); fdelplan@eso.org (F.D.-S.); gfinger@eso.org (G.F.); jkolb@eso.org (J.K.); emueller@eso.org (E.M.); lpallanc@eso.org (L.P.); jwoillez@eso.org (J.W.); gzins@eso.org (G.Z.)\\
$^{3}$ \quad Max-Planck-Institut f\"{u}r  Extraterrestrische Physik, 85748 Garching, Germany; soylentdeen@protonmail.com~(C.D); eisenhau@mpe.mpg.de (F.E.); eisenhau@mpe.mpg.de (G.C.) \\
$^{4}$ \quad {CNRS, IPAG,  Universit\'{e} Grenoble Alpes,}
 {38000 Grenoble}, France; Zoltan.Hubert@univ-grenoble-alpes.fr}

\corres{Correspondence: hippler@mpia.de}

\abstract{
This article describes the operation of the near-infrared wavefront sensing based Adaptive Optics (AO) system CIAO.
The Coud\'{e} Infrared Adaptive Optics (CIAO) system is a central auxiliary component 
of the Very Large Telescope (VLT) interferometer (VLTI).
It enables in particular the observations of the Galactic Center (GC) using the GRAVITY instrument. 
GRAVITY is a highly specialized beam combiner, a device that coherently combines the light of the 
four 8-m telescopes and finally records interferometric measurements in the K-band on 6 baselines simultaneously.
CIAO~compensates for phase disturbances caused by atmospheric turbulence, 
which all four 8\,m Unit Telescopes (UT) experience during observation. 
Each of the four CIAO units generates an almost diffraction-limited image quality at its UT, which ensures that maximum flux of the
observed stellar object enters the fibers of the GRAVITY beam combiner.
We present CIAO performance data obtained in the first 3 years of operation as a function of weather conditions. We describe how CIAO is configured and used for observations with GRAVITY. In addition, we focus on the outstanding features of the near-infrared sensitive Saphira detector, which is used for the first time on Paranal, and show how it works as a wavefront sensor detector.}


\begin{document}

\section{Introduction}
\label{gravIntro}  

What does a black hole look like? Do black holes really exist?
Questions like these can actually be answered more quantitatively thanks to better and better observations.
One of the most interesting observation areas in this context is the center of the Milky Way, 
where the black hole with the largest angular diameter as seen from Earth is located.
The compact radio source Sagittarius A* (SgrA*) at the very center of the Milky Way has been observed over more than five decades (see Genzel et al. \cite{Genzel2010} and references therein). 
In 2009, Gillessen et al. \cite{Gillessen2009} published a study which monitored stellar orbits around SgrA* for 16 years.
They estimate the mass of SgrA* at about 4 million solar masses. At the distance of the GC of $\approx$8\,kpc, the Schwarzschild radius $R_S$ for a black hole candidate with that mass is 10\,$\upmu$as, equivalent to a source diameter of $\approx$24 million km, or 0.16 astronomical units (AU). Very long baseline radio interferometric measurements, observing the---gravitational lensing magnified---black hole shadow in the emission of SgrA*, give an angular source width of 37\,$\upmu$as \cite{Doeleman2008}.
This width is actually smaller than the expected width of 5~$R_S = 50\,\upmu$as because of gravitational lensing.
The deviation can be explained using different models for the intrinsic structure of SgrA* \cite{Lu2018}.

Due to the extremely high extinction in the visible spectral range, the GC can only be examined from the ground at radio and infrared wavelengths. In order to achieve the required spatial resolution in the order of $R_S$, phase-referenced observations at the VLTI in the infrared spectral range are suitable~\cite{Woillez2018}.
The instrument to implement this is called GRAVITY \cite{GRAVITY2011} and was proposed in 2005 \cite{Eisenhauer2008}. GRAVITY had first light with all four 8\,m UTs in May 2016 \cite{GRAVITY1stlight2017}.

Recent observations with the GRAVITY instrument allowed for the first time to measure the gravitational redshift in the light of a star orbiting SgrA*
and approaching the GC to about 120~AU, i.e., 1400 $R_S$ \cite{GRAVITY2018a}.
Further results show that GRAVITY reaches an astrometric precision close to \mbox{10\,$\upmu$m~as~\cite{GRAVITY2018b, GRAVITY2018c}.}
In imaging mode, GRAVITY can achieve angular resolutions with the 4 UTs down to 4\,mas (milli-arcsecond).

For the same reasons that the infrared range was used to observe the GC with GRAVITY, the~visible AO system MACAO \cite{Schoeller2007}, already in operation at the VLTI since 2003, had to be supplemented by an infrared-based AO system. This addition, called CIAO, is discussed in more detail in the following~sections.

Although this report is mainly about the AO system performance during GC observations with GRAVITY, there are many other areas in which GRAVITY can be used. This ranges from interferometric Exoplanet observations to mapping the cores of active galactic nuclei \cite{Messenger2019}.

\section{Galactic Center Observations with GRAVITY: Requirements for CIAO}
\label{ciaoGC}

\begin{figure}[H]
\centering
\includegraphics[width=14cm]{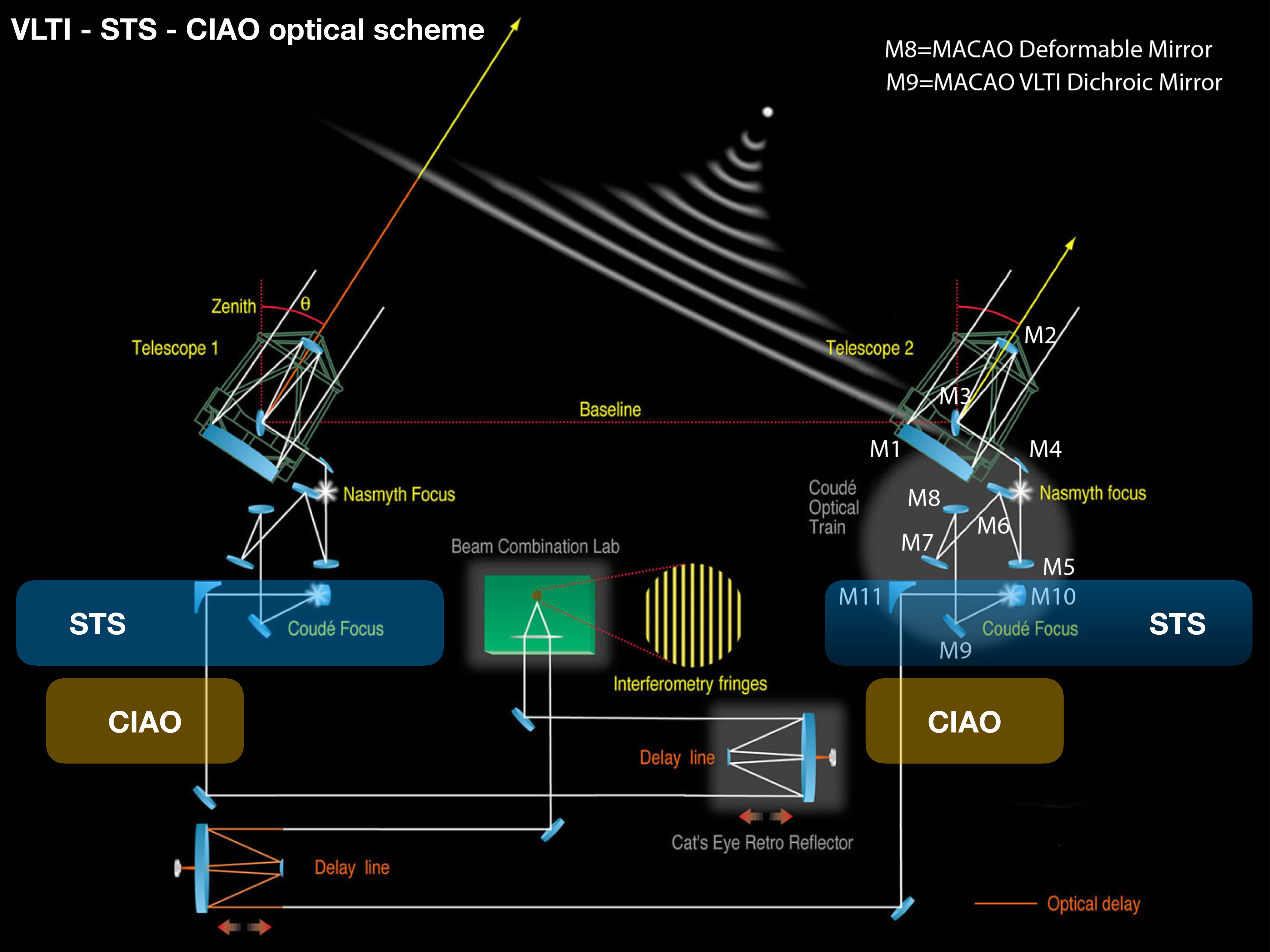}
\caption{\label{VLTIpath}Schematics of the Very Large Telescope (VLT) interferometer optical path with 2 Unit Telescopes (UTs). The CIAO units are located in the Coud\'{e} focal station of the respective telescope. Light from a celestial object is reflected by 9 mirrors (Telescope 2: M1--M9) and is in focus at the mirror M10, located in the Coud\'{e} focal station. Star separator units (STS) units select 2 celestial objects within a field of view of 2\,arcmin diameter. The light of one  object is relayed to CIAO, the light of the second object is relayed to the GRAVITY beam combiner, located in the beam combination lab. Delay lines compensate for optical path length differences between telescopes 1 and 2, which have a distance corresponding to the baseline. Mirror M8 is the deformable mirror shared between MACAO and CIAO. Mirror M9 is a dichroic beam-splitter that reflects infrared light to CIAO and the beam combination lab, visible light is transmitted to MACAO (not shown). Background image: European Southern Observatory (ESO).}
\end{figure}

GRAVITY is an interferometer that can combine the light of the 4 UTs such that all combinations of 2 UTs, six in total, create interferometric fringes. As described in detail in \cite{Yang2013, Pfuhl2014, Perraut2018}, after passing the delay lines of the VLTI (see also Figure~\ref{VLTIpath} and Section~\ref{CIAOoverview}) and the GRAVITY fiber coupler unit, the light of the 4 UTs is fed into single-mode fibers and then further relayed into an integrated optics unit where pairs of UT beams interfere. GRAVITY observes in the near-infrared (NIR) spectral range centered at 2.2~$\upmu$m (K-band, 1.9--2.45\,$\upmu$m). The GRAVITY single-mode fibers have a mode field diameter of 7.66\,$\upmu$m, sufficiently large to collect the image of a diffraction-limited 2.2\,$\upmu$m point source. The fiber coupler injection optics has an f-number of 2.5. This focuses the (nearly) diffraction-limited K-band point spread function (PSF) with a full width at half maximum (FWHM) of 5.5\,$\upmu$m into the fiber. 

The task of CIAO is to create exactly this (nearly) diffraction-limited K-band PSF at the entrance of the fiber coupler inside the GRAVITY beam combiner. The top-level requirements of GRAVITY for CIAO taken from \cite{Clenet2010} are listed in Table~\ref{CIAOperf_requirements}.

\begin{table}[H]
\centering \small
\caption{Performance requirements for Coud\'{e} Infrared Adaptive Optics (CIAO) under average atmospheric conditions with a median seeing value at $\lambda$ = 0.5 ${\upmu}$m and at zenith of 0.85\,arcsec.}
\label{CIAOperf_requirements}
\begin{tabular}{ccccc}
\toprule
\textbf{Guide Star }    & \textbf{ Guide Star Separation }        &  \textbf{ Telescope      } &  \textbf{ Strehl Number }                   & \textbf{Residual 2-Axis} \\
\textbf{ Magnitude}    &  \textbf{from Science Target  }          &  \textbf{ Zenith Angle }   &   \textbf{at }\boldmath{$\lambda$} \textbf{= 2.2} \boldmath{$\upmu$}\textbf{m}  & \textbf{Image Jitter} \\
 \textbf{   in K-Band } & \textbf{ [Arcsec]  }                             &   \textbf{ [Degrees]      } &    \textbf{[\%]                                  } &\textbf{ [mas rms]} \\
\midrule
7 \textsuperscript{a}      & 7       & 30        & 25  & 10 \\
7      & 0       & 30        & 35   & 10 \\
10   &  0       & 30        & 10  & 22 \\ 
\bottomrule
\end{tabular}
\begin{tabular}{ccccc}
\multicolumn{1}{c}{\footnotesize \textsuperscript{a} Galactic Centre supergiant IRS7-like case, see Figure~\ref{STSandGC}b.}
\end{tabular}
\end{table}
\vspace{-8pt}
The {\bf functional requirements} are:
\begin{itemize}
\item provide wavefront correction.
\item provide near infrared wavefront sensing---allow for off-axis wavefront  sensing.
\end{itemize}

All requirements could be fulfilled with a 9\,$\times$\,9 Shack-Hartmann wavefront sensor-based 
AO system running at a speed of 500\,Hz. 
A 320\,$\times$\,256 pixel HgCdTe avalanche photodiode (APD) array produced by Leonardo S.p.A. in the U.K. was selected as the infrared detector (see Section~\ref{SAPHIRA}).

\section{Overview of the CIAO System}
\label{CIAOoverview}

CIAO \cite{Scheithauer2016, Deen2016, Kendrew2012, Clenet2010, Hippler2008} was built to equip the VLTI instrument 
GRAVITY with infrared-sensitive adaptive optics. GRAVITY's primary scientific object of investigation is the center of the 
Milky Way and there especially the massive black hole.
The purpose of CIAO is to increase the sensitivity of GRAVITY. This~is achieved by compensating for atmospheric turbulence and the resulting almost diffraction-limited image quality of each individual UT. For the GRAVITY instrument, this means 
a stable injection of light from two adjacent astrophysical objects into the two single-mode input fibers per telescope of GRAVITY. While GRAVITY is located in the beam combination lab (see Figure~\ref{VLTIpath}), the four almost identical CIAO units are located inside the Coud\'{e} focal stations of the four UTs. Each CIAO unit was designed to fit seamlessly into the VLTI infrastructure.  An important element of the existing infrastructure, the~60\,actuators bimorph deformable mirror of the MACAO optical AO system, is~shared by CIAO.

\begin{figure}[H]
\centering
\subfigure[]{
 \label{STSpath}
\includegraphics[width=85mm]{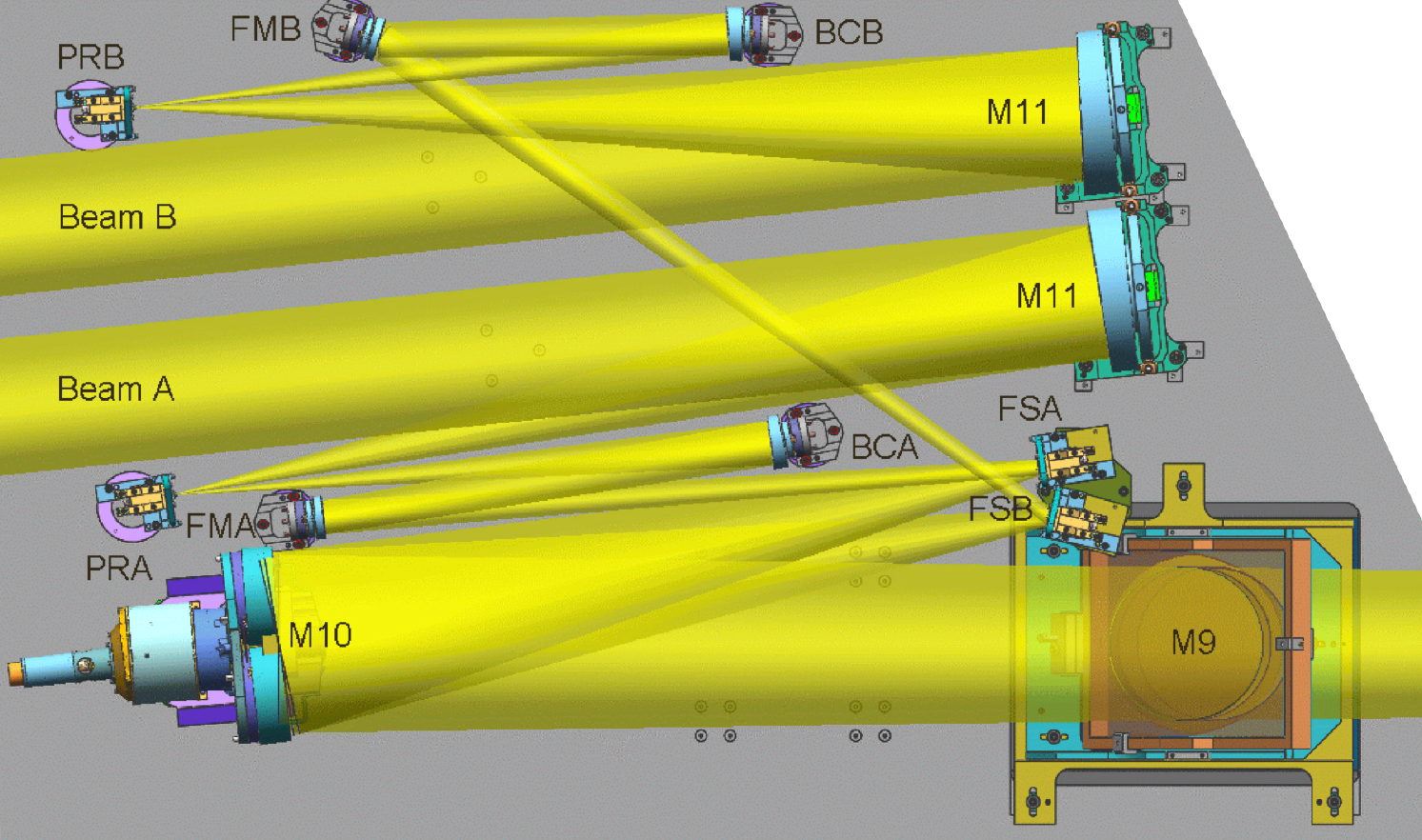}
}
\subfigure[]{
\label{GCsetup}
\includegraphics[width=58mm]{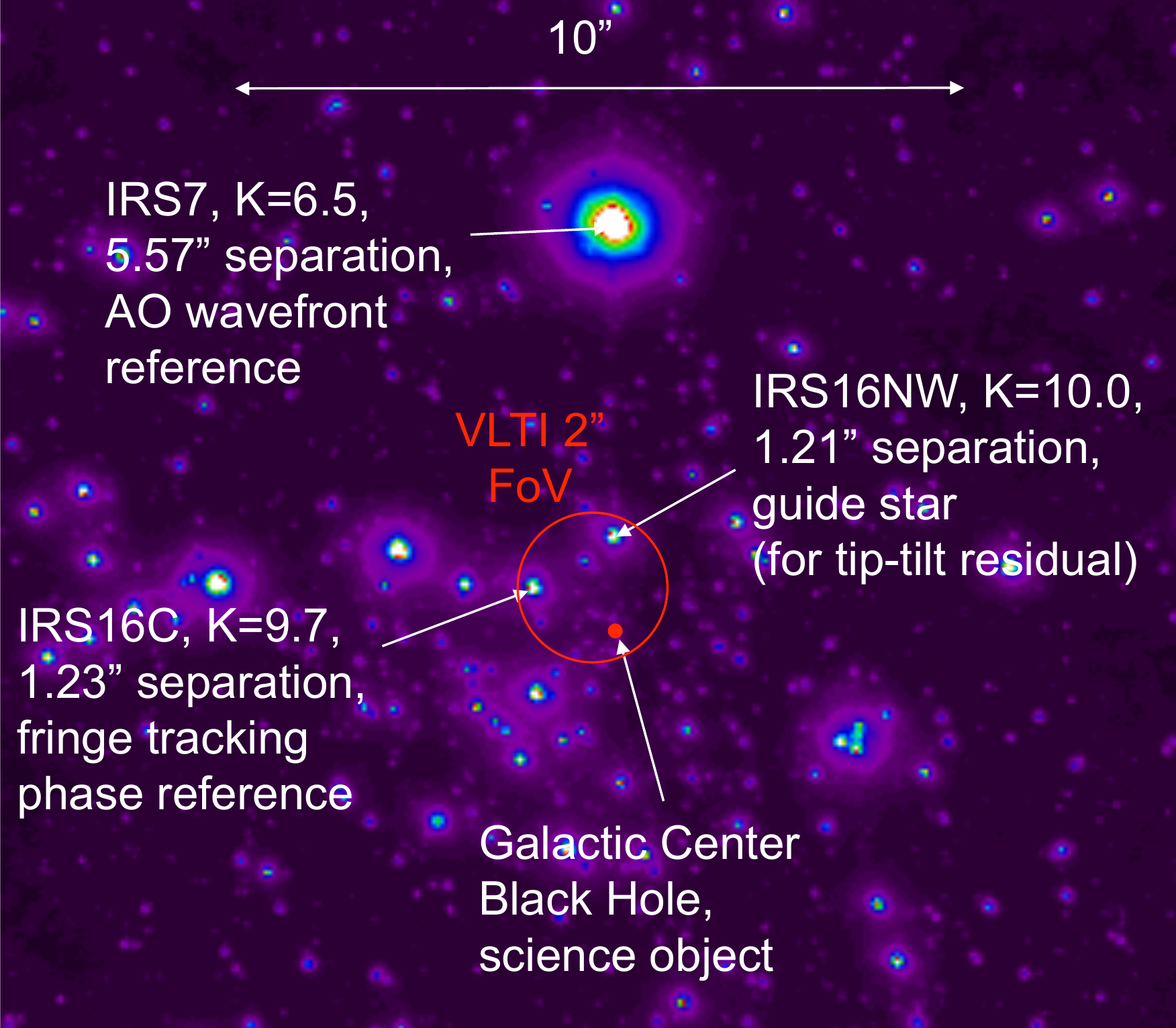}
}
\caption{STS optical layout and on-sky set-up for GC observations with GRAVITY. See text for details. (\textbf{a}) Seen perpendicular to the image plane, light from the telescope is reflected on the dichroic beam-splitter M9 towards two mirrors labeled M10. The star separator (STS) unit uses two field select mirrors labeled FSA and FSB to pick the light of two objects A and B, focused on the M10 pair. FSA and FSB are able to scan M10. 
The light of the two selected objects is then relayed via folding mirrors (labeled FMA and FMB) towards the output beams A and B. Before, however, the~beams are compressed via two elliptical mirrors labeled BCA and BCB, and then forwarded via the pupil re-imagers PRA and PRB. Each output beam contains a field of view of about 2$''$\, $\times$ \,2$''$. Image: ESO; (\textbf{b}) STS set-up for GC observations with GRAVITY. Light from the infrared source IRS7 is transmitted via the STS beam B to the CIAO wavefront sensor. Light from inside the VLTI 2$''$ field of view (red circle) is transmitted via the STS beam A to to GRAVITY beam-combiner. }
\label{STSandGC}
\end{figure}

Figure~\ref{VLTIpath} shows the VLTI optical scheme for the sake of simplicity for two UTs instead of all four UTs. Included are the locations of the STS. Their purpose is to select two small fields of the sky and feed them forward towards the VLTI beam combination lab \cite{Nijenhuis2005, Nijenhuis2008}. Details about the STS are shown in Figure~\ref{STSandGC}a. The standard set-up for GC observations is shown in Figure~\ref{STSandGC}b.

\begin{figure}[H]
\centering
\includegraphics[width=14cm]{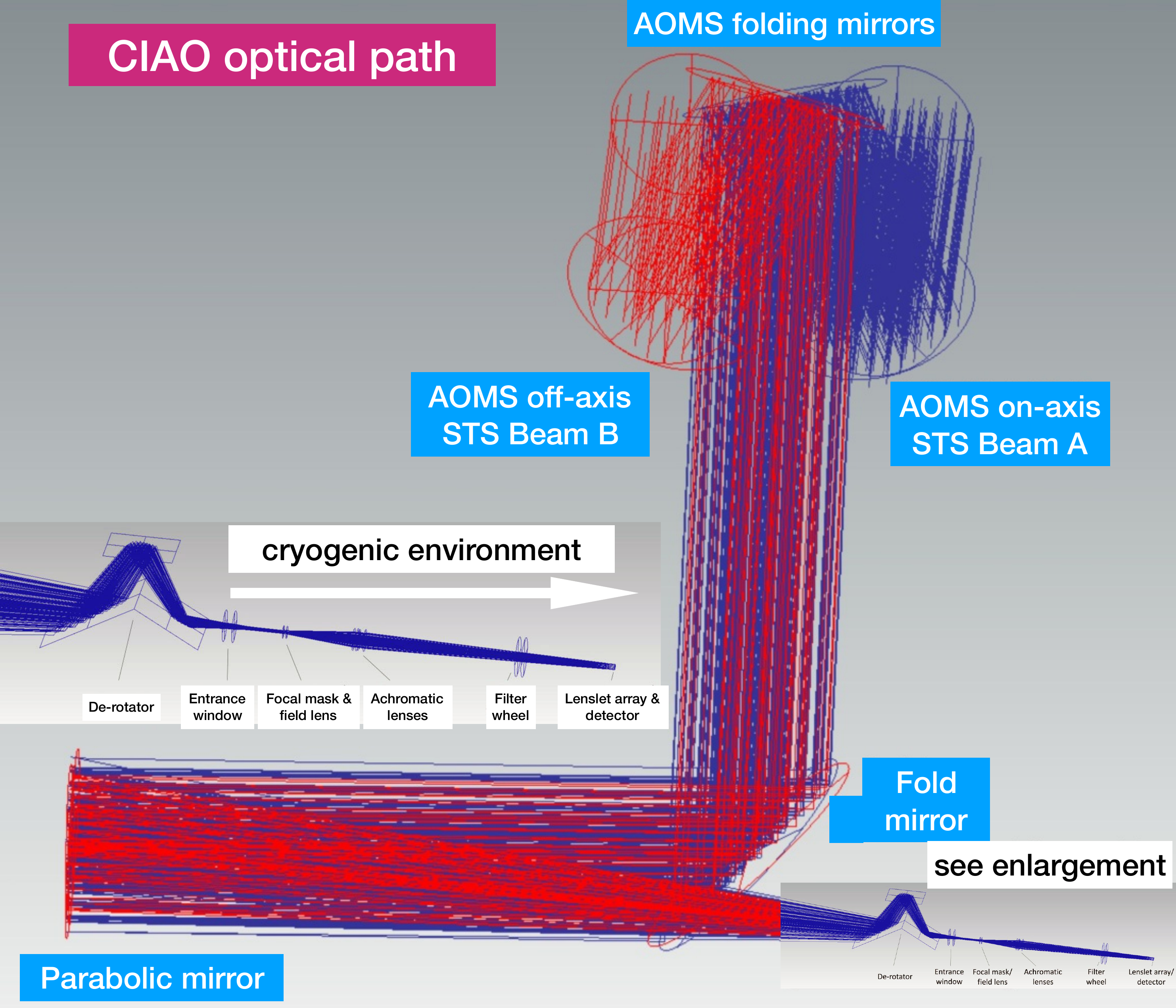}
\caption{\label{CIAOpath} Using the adaptive optics (AO) mode selector (AOMS), each CIAO unit can select either STS beam A or B as optical input channel. 
The light from the selected channel is then relayed via a fold mirror and a parabolic mirror through a de-rotator unit (K-mirror design) into the CIAO wavefront sensor cryostat. 
The cryogenic part on the lower right side is shown enlarged on the left side in the middle below the
label {cryogenic environment}.}
\end{figure}

The CIAO units are designed to deliver nearly diffraction-limited image quality in K-band for each UT. 
To achieve this, the light of a not too faint point source, i.e., the wavefront sensor (WFS) reference star, with m$_K\leq$ 10 is required. 
For GC observations the WFS reference star is the bright M-type supergiant IRS7, located about 5.6$''$ away from SgrA*.
In this case, the STS beam B points to IRS7 and CIAO picks this beam using its movable off-axis AO mode selector (AOMS) unit accordingly.
The light of the reference star is then relayed and focused via a fold mirror and a parabolic mirror to a focal plane inside the CIAO WFS cryostat (see Figure~\ref{CIAOpath}). Alternatively, the GRAVITY beam-combiner and CIAO can share the light of a source. In this case, STS beam A points to the science object and CIAO moves its AOMS on-axis unit to pick beam A. The difference between AOMS off-axis and AOMS on-axis is that the off-axis unit is fully reflective, while the on-axis unit is equipped with a beam splitter. The~beam splitter reflects light towards CIAO and transmits light towards the GRAVITY beam-combiner. In~the focal area of the CIAO cryostat there is a focal mask and a field lens. The~lateral movable field lens together with an achromatic doublet image the pupil (image of the MACAO deformable mirror) through a bandpass filter onto the lenslet array of the Shack-Hartmann sensor. 

The lenslet array, located in the pupil plane, 
focuses its images directly on the Saphira detector (see~the enlargement of the cryogenic environment in Figure~\ref{CIAOpath}).
The lateral movable field lens allows us to stabilize the registration between the lenslet array and the deformable mirror 
in closed-loop operation \cite{Dai2017}.
The lenslet array is optimized for a wavelength of 2.2\,$\upmu$m and a detector pixel size of 24\,$\upmu$m. It consists of 9\,$\times$\,9 square lenses in a square geometry. Each square lens has a size of 192\,$\times$\,192\,${\upmu}$m$^2$ and a focal length of 2095\,$\upmu$m for a wavelength of 2.2\,$\upmu$m. The diffraction-limited K-band PSF of a single lens has an FWHM of $\leq$26\,$\upmu$m, about the size of one detector pixel. 
The~CIAO lenslet array requires exactly 72\,$\times$\,72\,pixels on the detector. Due to the multi-channel readout electronics, CIAO~reads a rectangular area of 96\,$\times$\,72\,pixels on the detector. More details about the read-out modes of the Saphira detector are given in Section~\ref{SAPHIRA}.

Finally, the transmission of all optics up to detector at $\lambda$ = 2.2\,$\upmu$m for the two operating modes is 
T = 0.15 for the on-axis configuration and T = 0.32 for the off-axis configuration. 
These estimates include neither the quantum efficiency of the detector nor the atmospheric transmission.

\section{The SAPHIRA Detector}
\label{SAPHIRA}

\begin{figure}[H]
\centering
\subfigure[]{
\label{lensletMicroscope}
\includegraphics[width=72mm]{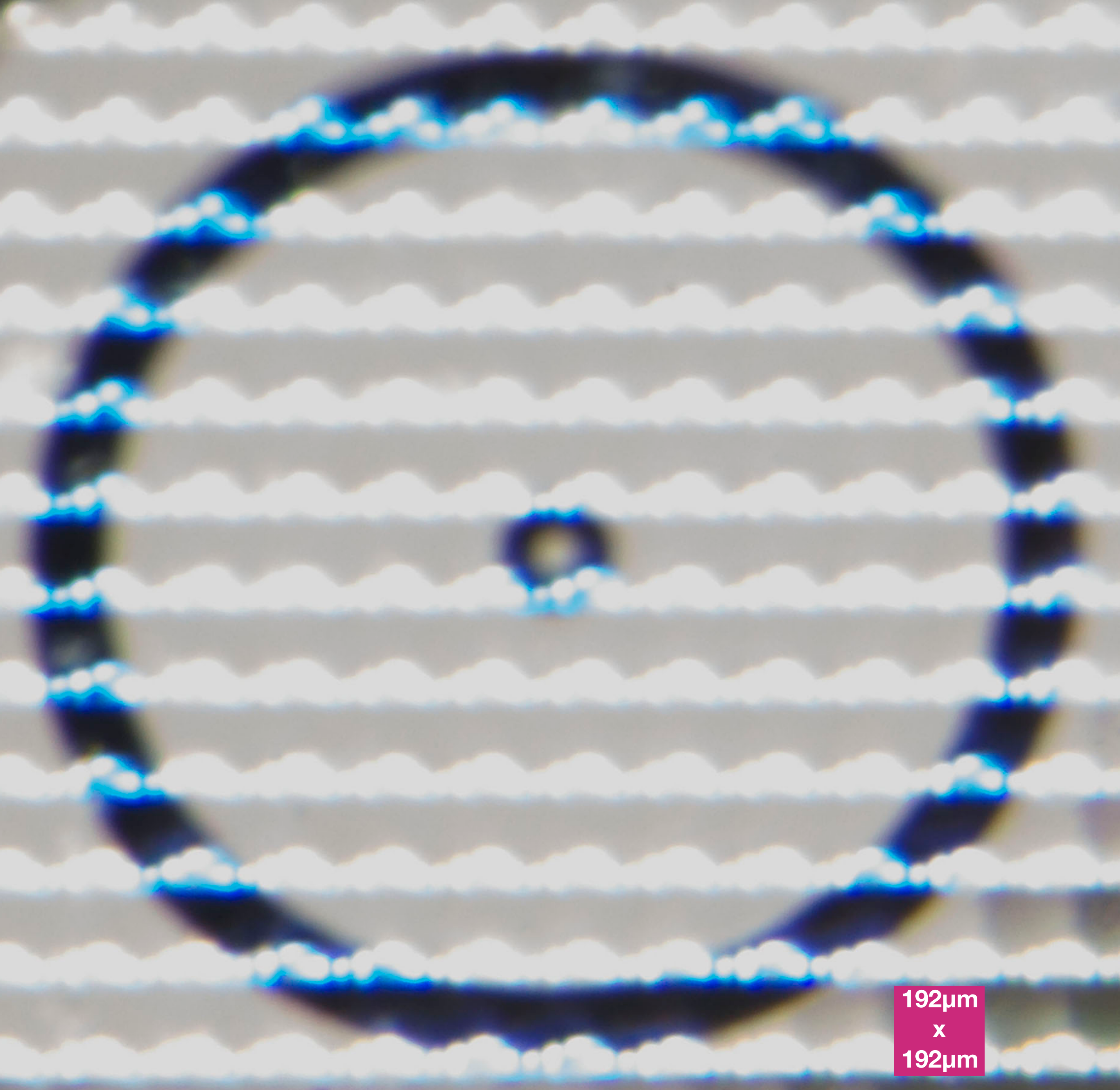}
}
\subfigure[]{
\label{lensletFootprint}
\includegraphics[width=70mm]{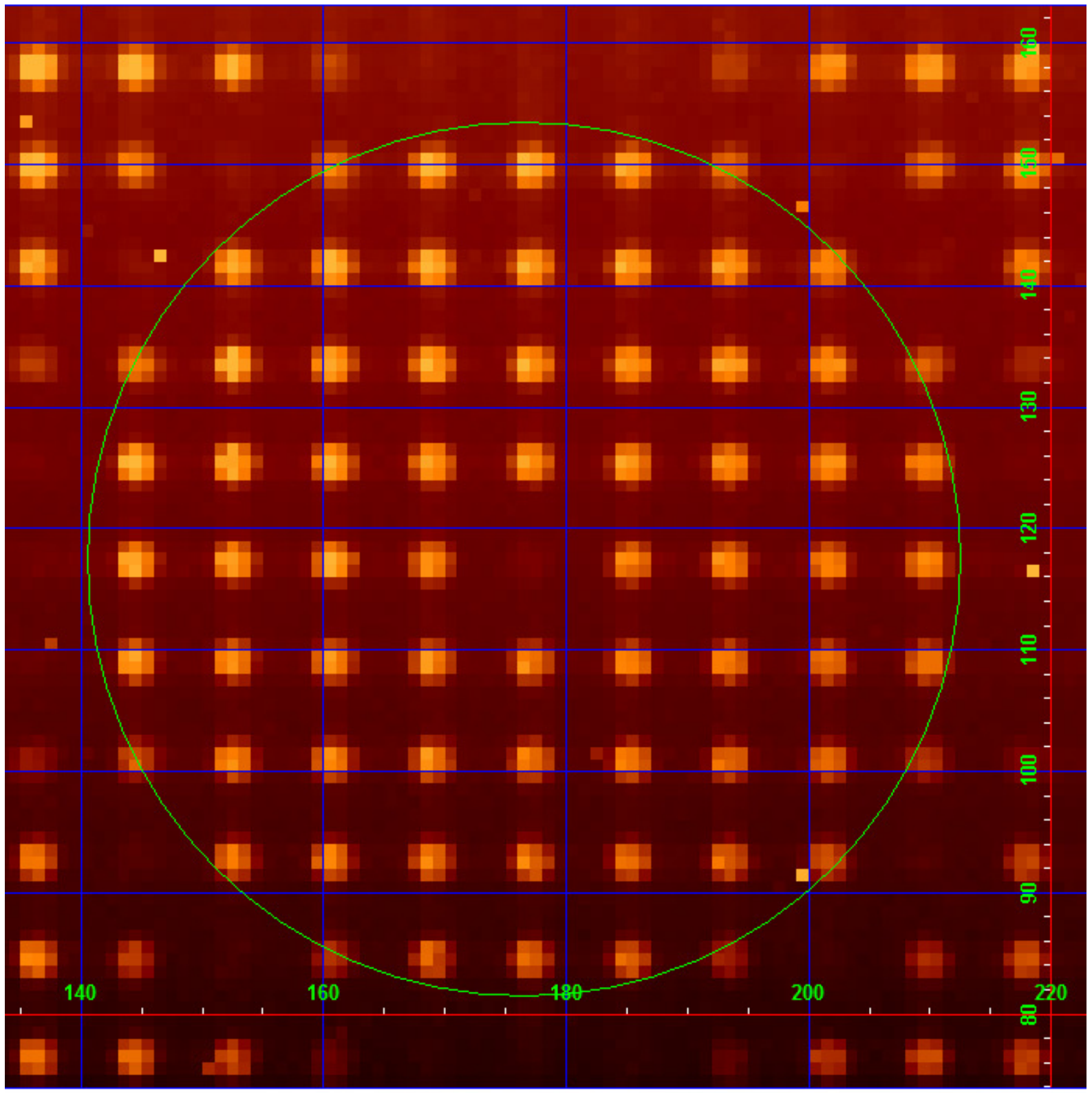}
}
\caption{\label{llaDesignFootprint}CIAO lenslet design and footprint on the detector. See text for details. (\textbf{a}) Microscopic image of a CIAO lenslet array with ``blue'' light reflecting chromium masks, defining the pupil boundaries.
                 The scale is indicated by the red square; (\textbf{b}) Footprint of the illuminated CIAO lenslet on the Saphira detector. Green circle diameter: 72\,pixels.}
\end{figure}

Due to their high frame rates, high sensitivity, low noise, and low dark current, Saphira
\mbox{detectors \cite{Finger2014, Finger2016, Mehrgan2016, Atkinson2016, Finger2019, Goebel2018}} have characteristic properties, 
especially for applications in astronomical adaptive optics, which were not available 10 years ago.
The Saphira detector used in CIAO is  a  \mbox{320\,$\times$\,256 pixel} HgCdTe avalanche photodiode array with 24\,$\upmu$m square pixels.
It is sensitive in the spectral range from 0.8--2.5\,$\upmu$m. Unlike other near-infrared arrays, Saphira
features  a  user-adjustable  avalanche  gain,  which~multiplies  the  photon  signal  but  has a 
minimal  impact  on  the  read  noise. The CIAO Shack--Hartmann lenslet array, a 9\,$\times$\,9 square design, is
located in the pupil plane such that 68 of the 81 sub-apertures are illuminated (Figure~\ref{llaDesignFootprint}). Each sub-aperture 
covers exactly 8\,$\times$\,8 pixels on the detector, a total area of 72\,$\times$\,72 pixels on the detector is evaluated for signal evaluation.

To realize an analog bandwidth for high frame rates, the CIAO Saphira detector array has 32~parallel outputs.
To further increase the frame rates of sub-regions, the 32 outputs of the readout integrated circuit (ROIC)
are organized to read 32 adjacent pixels of a row at the same time.

To fit the required CIAO lenslet detector region (72 $\times$ 72 pixels) with this parallel design, a~detector sub-region of 96\,$\times$\,72 = 6912\,pixels has to be readout.
Since the readout electronics reads each 320\,pixel long detector line in 10 channels of 32\,pixels each, 3 channels (96\,pixels) must be read out for each \mbox{72 CIAO pixels} along one line.

Pixel information is digitized using 5\,MHz analog to digital converters (ADC). 
Considering clocking overheads, the time for a non-destructive
readout of the CIAO detector sub-region takes 70.6\,$\upmu$s \cite{Finger2016}.

The maximum CIAO loop speed of 500\,Hz, therefore, allows us to oversample the detector sub-region up to 28 times using 
Fowler sampling techniques \cite{Fowler1990, Finger2016}.
In particular, this together with a proper electron APD (eAPD) gain setting significantly reduces effective detector readout-noise, 
down to below 1\,electron  per pixel. All 4 CIAO Saphira detectors are based on metal--organic vapor phase epitaxy (MOVPE), 
and have been produced by Leonardo S.p.A. (formerly Selex ES) \cite{Leonardo2016}. The CIAO Saphira detectors are so-called Mark3 devices. It should be noted that the GRAVITY fringe tracker, which compensates the differential piston based on measurements of a brighter off-axis astronomical reference source, uses a Saphira detector as well \cite{Lacour2019}. 
Table~\ref{saphiraData} summarizes the most important CIAO detector characteristics. 

\begin{table}[H]
\centering \small
\caption{CIAO Saphira detector characteristics @ 95 K temperature, or as indicated}
\label{saphiraData}
\begin{tabular}{lccccr}
\toprule
\textbf{Saphira Detector Number}                                & \textbf{ \#1 }       &   \textbf{\#2}        &   \textbf{\#3}            &  \textbf{\#4 }      & \textbf{Comment} \\
\midrule
Readout noise [e$^-$ rms/pixel ]$^{\text{a}}$ & 1.24       & 0.95        & 1.17             & 0.96      & Fowler-8$^{\text{b}}$  \\
Bad pixels [pixel/detector]                              & 63           & 39           & 78                & 40         & CDS$^{\text{c}}$ \\
Bad pixel in CIAO sub-region                         &  3            &  1            &  0                 &  2           & \\ 
Detector gain [e$^-$/ADU]                              & 0.380      & 0.387      & 0.388         & 0.395     & CDS$^{\text{c}}$ \\
eAPD gain [e$^-$/e$^-$]                                 & 30.11      & 30.64      & 26.40         & 27.15     & \\
Dark current  [e$^-$/pixel/s]                            & 478.6      & 485.15    & 631.1         & 630.85  & CDS$^{\text{c}}$ \\
Quantum efficiency@90 K [e$^-$/$\gamma$] & \multicolumn{4}{c}{60\%, 67\%}                      & H-band, K-band \\
\bottomrule
\end{tabular}\\
\begin{tabular}{lccccr}
\multicolumn{1}{p{\textwidth -.88in}}{\footnotesize \textsuperscript{a} This is the effective readout noise, i.e., measured readout noise divided by eAPD gain; \textsuperscript{b} Measured using 8 Fowler pairs sampling combined with subpixel-4 sampling; \textsuperscript{c} Measured using correlated double sampling (CDS), i.e., 1 Fowler pair.}
\end{tabular}
\end{table}
\vspace{-6pt}

We want to emphasize that the readout method is also decisive for the noise behavior of the detector.
The CIAO SAPHIRA readout integrated circuit (ROIC) operates the detector on a frame by frame basis, i.e., the complete
frame is reset and then at least two frames are readout to obtain a so-called correlated double sampled (CDS) frame.
Finally, the two frames are subtracted from each other \cite{Finger2016}. 
This is a well-known technique to reduce various noise sources, including kTC noise, 
and~was already established for CCD detectors readout in the early 1970s \cite{White1974}.
Unfortunately, with the installed ROIC of the CIAO wavefront sensors, the duty cycle is only 50\%. 
A newer version of the ROIC, which can readout and reset the detector on a row by row basis (so-called rolling shutter mode), 
no longer has this problem \cite{Finger2016, Atkinson2016, Finger2019}.

Additionally, to further reduce readout noise, each pixel can be multiple times converted to digital values and then averaged.
As already mentioned, the CIAO window on the detector can be readout 28 times during exposure with an
integration time of 2\,ms. The point here is that this happens without resetting the pixels. This readout mode is usually called
non-destructive readout. In the end, oversampling and subtracting the pixels results in a significant reduction of the readout noise.
Such~readout schemes are nowadays also used for CCD detectors \cite{Alessandri2016, CruzDeLaTorre2018}.

\section{Statistical Review of Three Years of CIAO Operation}

In this section, we show the performance results of the CIAO system obtained for GC observations in the years 2017 to 2019.
Using the configuration as shown in Figure~\ref{STSandGC}b, the results can be compared quite well, since the same reference star, IRS7, was always observed by CIAO.
IRS7 has a K-band magnitude of K = 6.5, which at an AO loop frequency of 500\,Hz results in a WFS 
sub-aperture detector signal per frame of about 1000 photo-electrons. 
Taking into account the very low readout noise of the detector of about 1 photo-electron per pixel, 
CIAO can be operated with this configuration in a regime with a high signal-to-noise ratio.

To estimate seeing, recorded time series of 
CIAO Shack-Hartmann slopes, and DM commands are analyzed during closed-loop operation.
The procedure is based on analyzing 60 Zernike coefficients as well as DM command vectors.
The Zernike coefficients are created from expanding the Shack-Hartmann slopes into Zernike polynomials.
The temporal power spectral density of the Zernike coefficients, as well as the applied DM voltages (commands), were then used to
calculate the atmospheric seeing and the pseudo-Strehl numbers 
(see e.g., Fusco et al., 2004 \cite{Fusco2004}, \mbox{Snyder et al., 2016 \cite{Snyder2016}). }

\begin{figure}[H]
\centering
\subfigure[]{
\label{SVS}
\includegraphics[width=69mm]{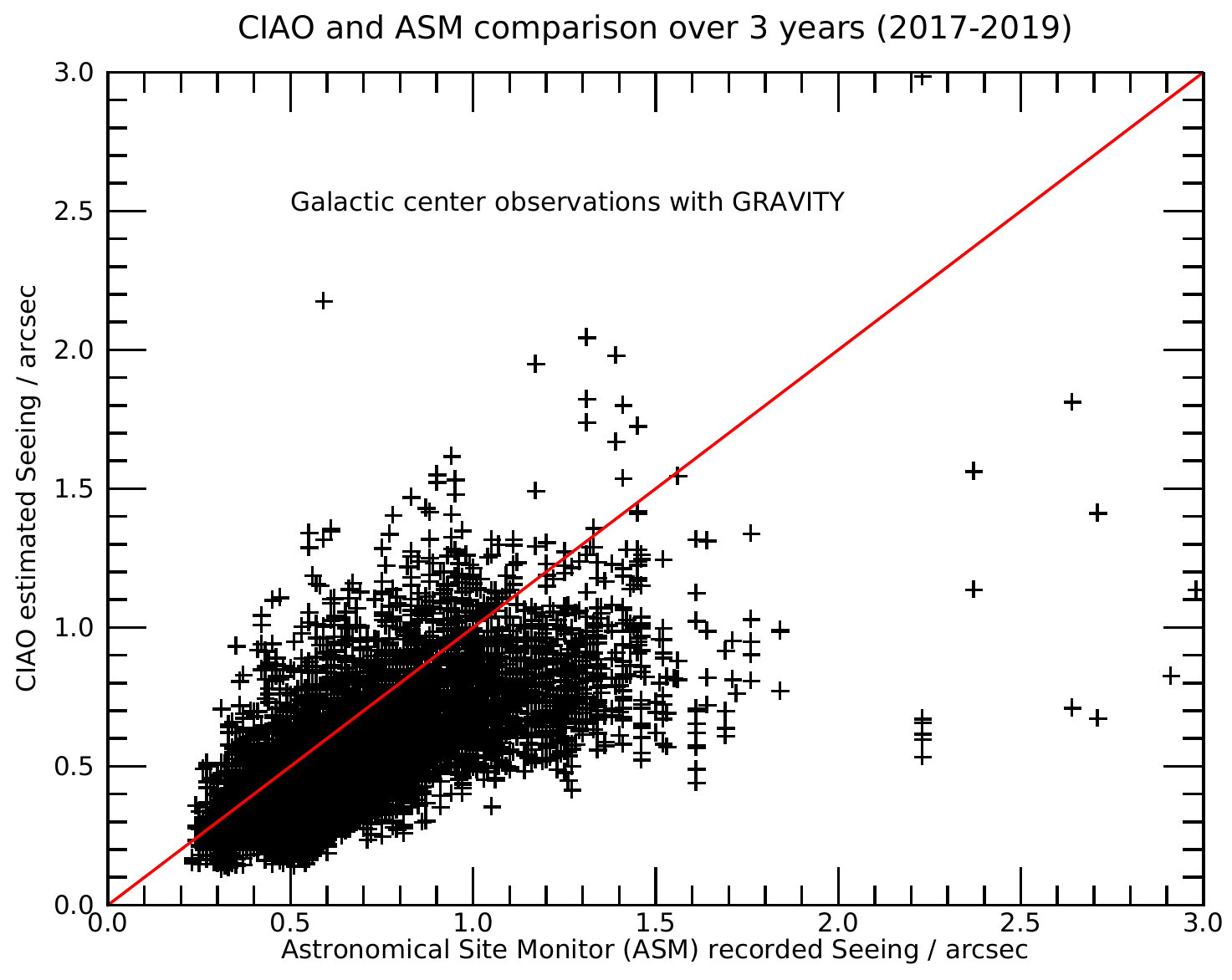}
}
\subfigure[]{
\label{CIAOperf}
\includegraphics[width=69mm]{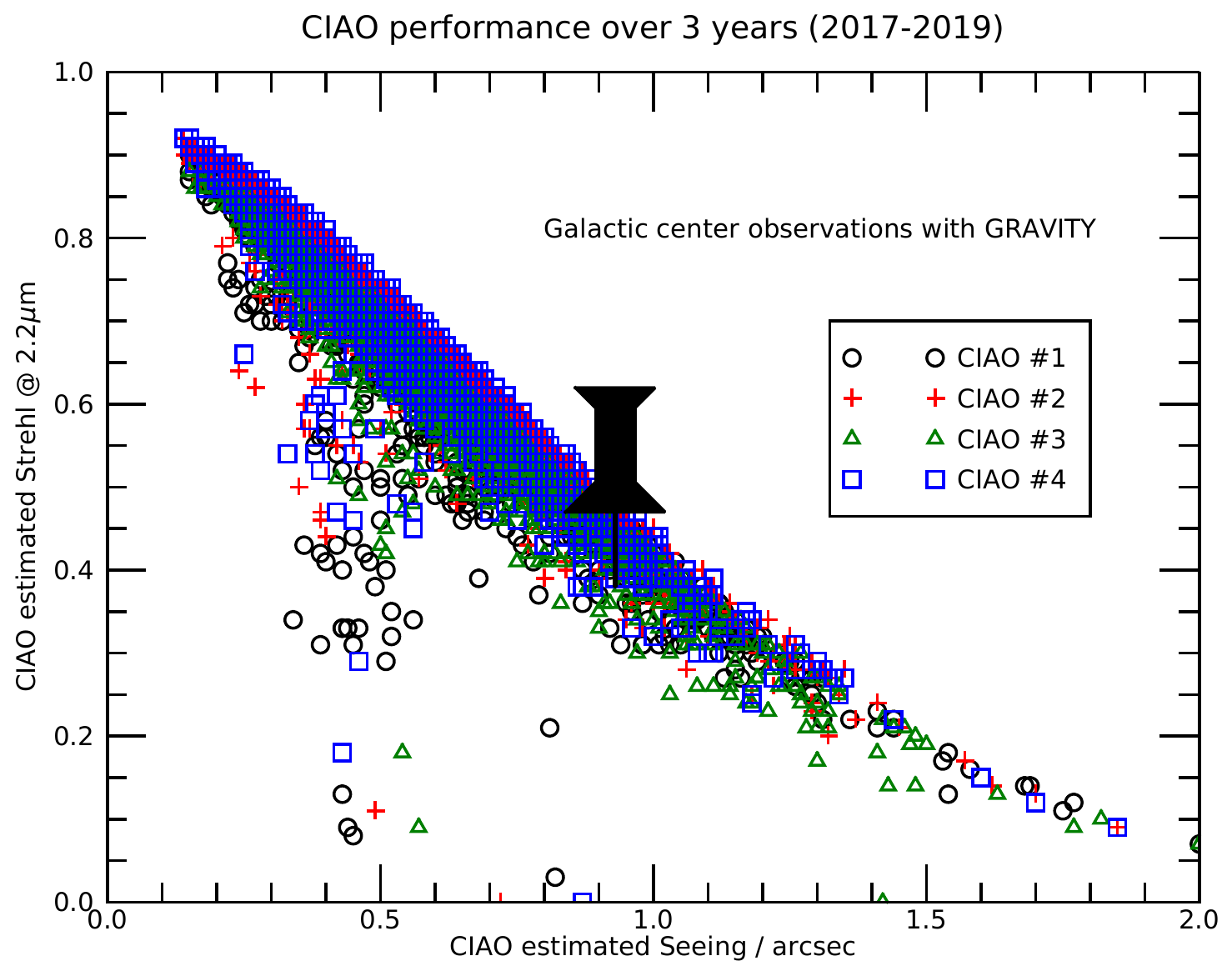}
}
\caption{Estimated CIAO seeing and CIAO performance results for 3 years of Galactic Center (GC) observations with GRAVITY.  (\textbf{a}) CIAO estimated seeing vs recorded Paranal Astronomical Site Monitor (ASM) seeing; (\textbf{b}) CIAO estimated K-band Strehl number vs CIAO estimated seeing (at $\lambda=0.5$ ${\upmu}$m 
and air mass along the line of sight). The pushpin points to our simulation result.}
\label{CIAOstats}
\end{figure}

Figure~\ref{CIAOstats}a compares the CIAO calculated seeing with the seeing recorded by the 
Paranal Astronomical Site Monitor (ASM) at the time of the CIAO observation.
The seeing values are calculated for zenith and for a wavelength of 500\,nm.
Please note that ASM and CIAO point to different
parts on the sky, and thus also different areas of the atmosphere. 
Figure~\ref{CIAOstats}b shows for each CIAO unit the estimated K-band ($\lambda=2.2$ ${\upmu}$m) Strehl as a function of the estimated seeing for Galactic Center observations in the years 2017 to 2019. This can be compared to our simulation result for GC observations \cite{Clenet2010, Kendrew2012}, which calculated a K-band Strehl of 0.38 at a zenith distance of 30\,degrees, and a seeing (at $\lambda=0.5$ ${\upmu}$m) of 0.93\,arcsec. See pushpin in Figure~\ref{CIAOstats}b. In general, the results obtained are slightly better than expected from our simulation.

Furthermore, there are some ``failures'' visible, which are usually based on technical problems and sometimes also on wrong software settings. 
Although, in most of these cases an improvement was still achieved (note that the natural Strehl at $\lambda = 2.2$ ${\upmu}$m even for an excellent Seeing of 0.5$''$ is only approx. 0.02).
Without AO, on average less than 1\% of the stellar flux can be coupled into the single-mode fiber of the GRAVITY beam combiner.

\begin{figure}[H]
\centering
\includegraphics[width=141mm]{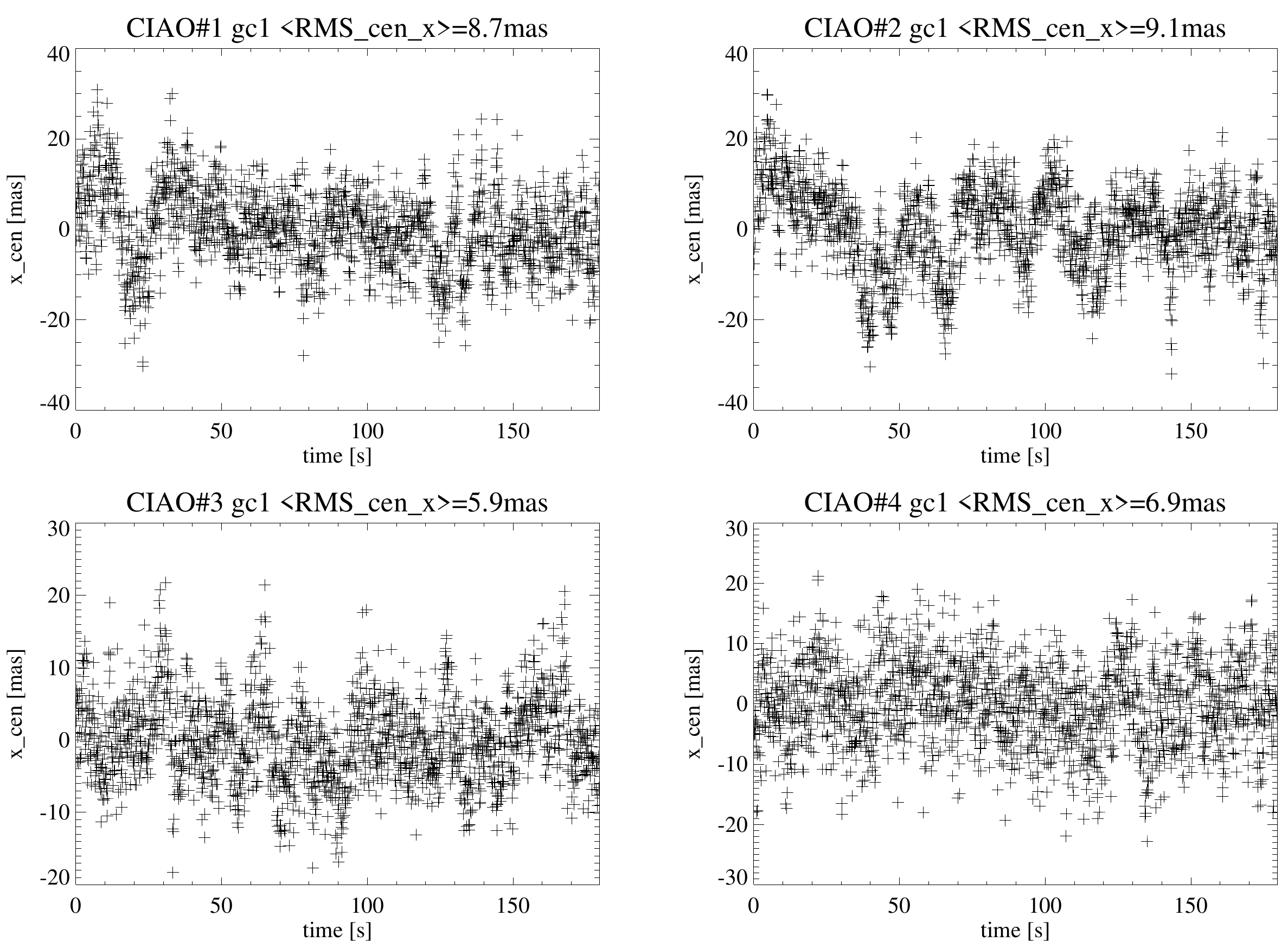}
\caption{\label{xjitter}Residual image motion jitter along the x-axis (x\_cen) over 150\,s observation of the GC. 
The~title line shows the calculated rms value (RMS\_cen\_x) in mas.}
\end{figure}

\begin{figure}[H]
\centering
\includegraphics[width=141mm]{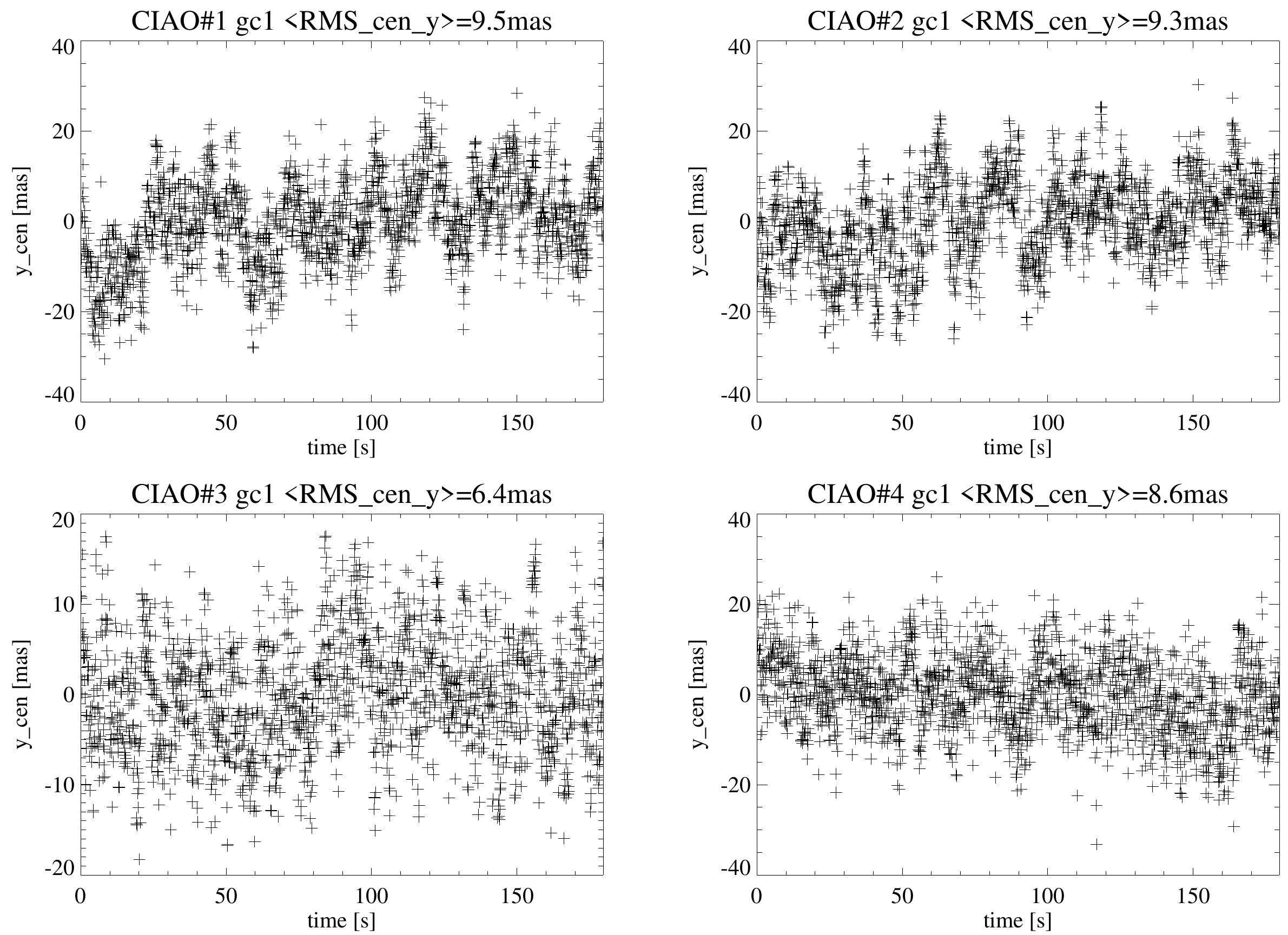}
\caption{\label{yjitter}Residual image motion jitter along the y-axis (y\_cen) over 150\,s observation of the GC. 
The~title line shows the calculated rms value (RMS\_cen\_y) in mas.}
\end{figure}

Besides the pure Strehl performance, the residual image jitter is an important factor to judge the efficiency of feeding light into the beam combiner fibers.
Figures~\ref{xjitter} and \ref{yjitter} show exemplarily for one night the residual image motion (jitter) over a period of 150\,s of all 4 CIAO units obtained for GC observations. In general, we find that the 1-axis residual image motion jitter is in the range \mbox{5--10\,mas~rms.}
This is well within the targeted range according to Table~\ref{CIAOperf_requirements}.
These measurements were taken with the VLTI infrared camera (IRIS) and 70\,ms integration time. 
IRIS observed a field around the GC, and one star of the IRS16 sources was used for the analysis.

\begin{figure}[H]
\centering
\includegraphics[width=124mm]{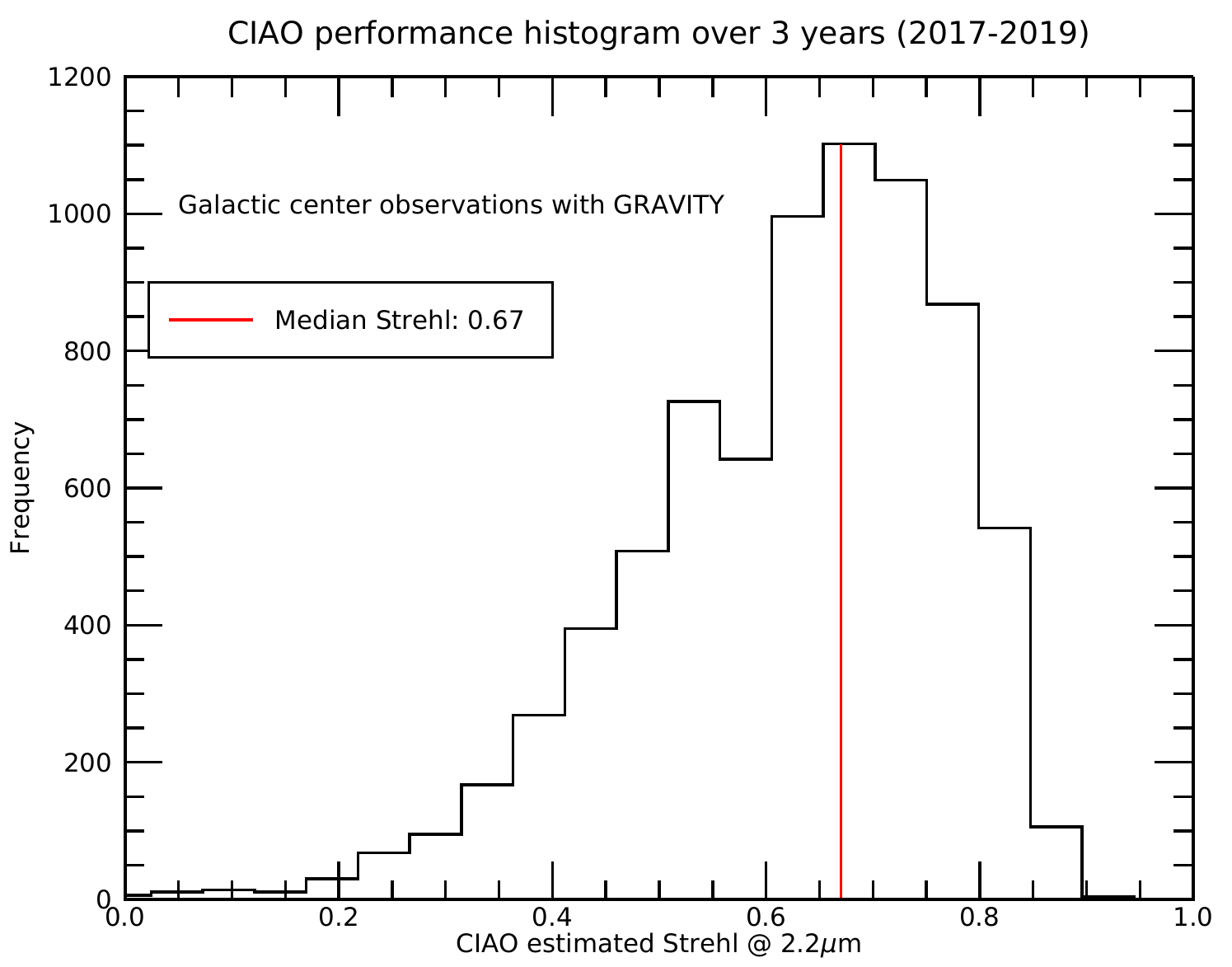}
\caption{\label{CIAOhistogram}CIAO Strehl histogram for 3 years of GC observations with GRAVITY. }
\end{figure}

As shown in Figure~\ref{CIAOhistogram},  GC observers had on average a very respectable AO performance in the years 2017--2019.
The achieved median Strehl number in K-band was above 0.6 and clearly met the Table~\ref{CIAOperf_requirements} requirement.

\unskip
\unskip
\unskip

\newpage
\section{Summary}
CIAO is an adaptive optics system, which in combination with the GRAVITY instrument, 
allows efficient interferometric observations with four 8~m telescopes.
In the three years studied, typically all CIAO units achieved a median K-band Strehl number of 0.67.
The first years of observations with CIAO and GRAVITY at the Paranal Observatory were very successful.
More than 30 scientific papers have been published by the GRAVITY collaboration since 2017. 
We anticipate that this will continue.
\vspace{6pt}

\authorcontributions{%
conceptualization, S.H. and W.B.; 
methodology S.H. and W.B.; 
project administration S.S.;
investigation, S.H., W.B., S.S., M.K., J.P., P.B., H.B., C.D., F.D.-S., F.E., G.F., Z.H., J.K.,
E.M., L.P., J.W., G.Z. and GRAVITY Collaboration; 
writing---review and editing, S.H. and W.B. All authors have read and agreed to the published version of the~manuscript.}

\funding{This research received no external funding.}
\conflictsofinterest{The authors declare no conflict of interest.}
\reftitle{References}
%
%

%
%
%

\begin{thebibliography}{999}
\providecommand{\natexlab}[1]{#1}

\bibitem[{Genzel} \em{et~al.}(2010){Genzel}, {Eisenhauer}, and
  {Gillessen}]{Genzel2010}
{Genzel}, R.; {Eisenhauer}, F.; {Gillessen}, S.
\newblock {The Galactic Center massive black hole and nuclear star cluster}.
\newblock {\em Rev.~Mod.~Phys.} {\bf 2010}, {\em 82},~3121--3195,
\newblock
  doi:{\changeurlcolor{black}\href{https://doi.org/10.1103/RevModPhys.82.3121}{\detokenize{10.1103/RevModPhys.82.3121}}}.

\bibitem[{Gillessen} \em{et~al.}(2009){Gillessen} et~al.]{Gillessen2009}
{Gillessen}, S.; Eisenhauer, F.; Trippe, S.; Alexander, T.; Genzel, R.; Martins, F.; Ott, T.
\newblock Monitoring Stellar Orbits Around the Massive Black Hole in the
  Galactic Center.
\newblock {\em Astrophys. J.} {\bf 2009}, {\em 692},~1075--1109,
\newblock
  doi:{\changeurlcolor{black}\href{https://doi.org/10.1088/0004-637X/692/2/1075}{\detokenize{10.1088/0004-637X/692/2/1075}}}.

\bibitem[{Doeleman} \em{et~al.}(2008){Doeleman} et~al.]{Doeleman2008}
{Doeleman}, S.S.; Weintroub, J.; Rogers, A.E.; Plambeck, R.; Freund, R.; Tilanus, R.P.; Friberg, P.; Ziurys, L.M.; Moran, J.M.; Corey, B.; et al.
\newblock {Event-horizon-scale structure in the supermassive black hole
  candidate at the Galactic Centre}.
\newblock {\em Nature} {\bf 2008}, {\em 455},~78--80,
\newblock
  doi:{\changeurlcolor{black}\href{https://doi.org/10.1038/nature07245}{\detokenize{10.1038/nature07245}}}.

\bibitem[{Lu} \em{et~al.}(2018){Lu} et~al.]{Lu2018}
{Lu}, R.S.; Krichbaum, T.P.; Roy, A.L.; Fish, V.L.; Doeleman, S.S.; Johnson, M.D.; Akiyama, K.; Psaltis, D.; Alef,~W.; Asada, K.; et al.
\newblock {Detection of Intrinsic Source Structure at ~3 Schwarzschild Radii
  with Millimeter-VLBI Observations of SAGITTARIUS A*}.
\newblock {\em Astrophys. J.} {\bf 2018}, {\em 859},~60,
\newblock
  doi:{\changeurlcolor{black}\href{https://doi.org/10.3847/1538-4357/aabe2e}{\detokenize{10.3847/1538-4357/aabe2e}}}.

\bibitem[{Woillez} \em{et~al.}(2018){Woillez} et~al.]{Woillez2018}
{Woillez}, J.; {Darr{\'e}}, P.; {Egner}, S.; {Gont{\'e}}, F.;
	{Haubois}, X.; {M{\'e}rand}, A.; {Schuhler}, N.; {Abad}, J.~A.;
	{Abuter}, R.; {Aller-Carpentier}; et al.
\newblock VLTI status update: Three years into the second generation.
\newblock  \emph{Proc.~SPIE}  \textbf{2018}, \emph{10701},~1070103.
\newblock
  doi:{\changeurlcolor{black}\href{https://doi.org/10.1117/12.2312042}{\detokenize{10.1117/12.2312042}}}.

\bibitem[{Eisenhauer} \em{et~al.}(2011){Eisenhauer} et~al.]{GRAVITY2011}
{Eisenhauer}, F.; Perrin, G.; Brandner, W.; Straubmeier, C.; Perraut, K.; Amorim, A.; Schöller, M.; Gillessen, S.; Kervella, P.; Benisty, M.; et al.
\newblock {GRAVITY: Observing the Universe in Motion}.
\newblock {\em Messenger} {\bf 2011}, {\em 143},~16--24.

\bibitem[{Eisenhauer} \em{et~al.}(2008){Eisenhauer} et~al.]{Eisenhauer2008}
{Eisenhauer}, F.; Perrin, G.; Rabien, S.; Eckart, A.; Lena, P.; Genzel, R.; Abuter, R.; Paumard, T.; Brandner, W.
\newblock {GRAVITY: The AO-Assisted, Two-Object Beam-Combiner Instrument for
  the VLTI}.
\newblock  In \emph{The Power of Optical/IR Interferometry: Recent Scientific Results
  and 2nd Generation}; {Richichi}, A., {Delplancke}, F., {Paresce}, F.,
  \mbox{{Chelli}, A., Eds.,}  {Springer: Berlin/Heidelberg, Germany,  }{2008}; p. 431.
\newblock
  doi:{\changeurlcolor{black}\href{https://doi.org/10.1007/978-3-540-74256-2_50}{\detokenize{10.1007/978-3-540-74256-2_50}}}.

\bibitem[{Gravity Collaboration} \em{et~al.}(2017){Gravity Collaboration}
  et~al.]{GRAVITY1stlight2017}
{Gravity Collaboration}; Accardo, M.; Amorim, A.; Anugu, N.; Avila, G.; Azouaoui, N.; Benisty, M.; Berger,~J.P.; Blind, N.; Bonnet, H.; et al.
\newblock First light for GRAVITY: Phase referencing optical interferometry for
  the Very Large Telescope Interferometer.
\newblock {\em Astron. Astrophys.} {\bf 2017}, {\em 602},~A94,
\newblock
  doi:{\changeurlcolor{black}\href{https://doi.org/10.1051/0004-6361/201730838}{\detokenize{10.1051/0004-6361/201730838}}}.

\bibitem[{Gravity Collaboration} \em{et~al.}(2018{\natexlab{a}}){Gravity
  Collaboration} et~al.]{GRAVITY2018a}
{Gravity Collaboration};  {Abuter}, R.; {Amorim}, A.; 
	{Anugu}, N.; {Baub{\"o}ck}, M.; {Benisty}, M.; {Berger}, J.P.;
	{Blind}, N.; {Bonnet}, H.; {Brandner}, W.; et al.
\newblock Detection of the gravitational redshift in the orbit of the star S2
  near the Galactic centre massive black hole.
\newblock {\em Astron. Astrophys.} {\bf 2018}, {\em 615},~L15,
\newblock
  doi:{\changeurlcolor{black}\href{https://doi.org/10.1051/0004-6361/201833718}{\detokenize{10.1051/0004-6361/201833718}}}.

\bibitem[{Gravity Collaboration} \em{et~al.}(2018{\natexlab{b}}){Gravity
  Collaboration} et~al.]{GRAVITY2018b}
{Gravity Collaboration}; {Abuter}, R.; {Amorim}, A.;
	{Baub{\"o}ck}, M.; {Berger}, J.P.; {Bonnet}, H.; {Brandner}, W.;
	{Cl{\'e}net}, Y.;  {Coud{\'e} Du Foresto}, V.; {de Zeeuw}, P.T.;
	  et al. 
\newblock {Detection of orbital motions near the last stable circular orbit of
  the massive black hole SgrA*}.
\newblock {\em Astron. Astrophys.} {\bf 2018}, {\em 618},~L10,
\newblock
  doi:{\changeurlcolor{black}\href{https://doi.org/10.1051/0004-6361/201834294}{\detokenize{10.1051/0004-6361/201834294}}}.

\bibitem[{Gravity Collaboration} \em{et~al.}(2018{\natexlab{c}}){Gravity
  Collaboration} et~al.]{GRAVITY2018c}
{Gravity Collaboration}; {Abuter}, R.; {Amorim}, A.;
	{Baub{\"o}ck}, M.; {Berger}, J.P.; {Bonnet}, H.; {Brandner}, W.; 
	{Cl{\'e}net}, Y.; {Coud{\'e} Du Foresto}, V.; {de Zeeuw}, P.T.; et al.
\newblock {Spatially resolved rotation of the broad-line region of a quasar at
  sub-parsec scale}.
\newblock {\em Nature} {\bf 2018}, {\em 563},~657--660,
\newblock
  doi:{\changeurlcolor{black}\href{https://doi.org/10.1038/s41586-018-0731-9}{\detokenize{10.1038/s41586-018-0731-9}}}.

\bibitem[Sch\"{o}ller(2007)]{Schoeller2007}
Sch\"{o}ller, M.
\newblock The Very Large Telescope Interferometer: Current facility and
  prospects.
\newblock {\em New~Astron.~Rev.} {\bf 2007},~{\em 51}, 628--638.
\newblock   doi:{\changeurlcolor{black}\href{https://doi.org/10.1016/j.newar.2007.06.008}{\detokenize{10.1016/j.newar.2007.06.008}}}.

\bibitem[{GRAVITY Collaboration} \em{et~al.}(2019){GRAVITY Collaboration}
  et~al.]{Messenger2019}
{GRAVITY Collaboration}; {Abuter}, R.; {Accardo}, M.;
         {Adler}, T.; {Amorim}, A.; {Anugu}, N.; {{\'A}vila}, G.;
         {Baub{\"o}ck}, M.; {Benisty}, M.; {Berger}, J.-P.; et al.
\newblock {GRAVITY Science}.
\newblock {\em   Messenger} {\bf 2019}, {\em 178},~19--49.
\newblock
  doi:{\changeurlcolor{black}\href{https://doi.org/10.18727/0722-6691/5170}{\detokenize{10.18727/0722-6691/5170}}}.

\bibitem[{Yang} \em{et~al.}(2013){Yang} et~al.]{Yang2013}
{Yang}, P.; Hippler, S.; Deen, C.P.; Brandner, W.; Clénet, Y.; Henning, T.; Huber, A.; Kendrew, S.; Lenzen, R.; Pfuhl, O.; et al.
\newblock {Characterization of the transmitted near-infrared wavefront error
  for the GRAVITY/VLTI Coud{\'e} Infrared Adaptive Optics System}.
\newblock {\em Opt. Express} {\bf 2013}, {\em 21},~9069.
\newblock
  doi:{\changeurlcolor{black}\href{https://doi.org/10.1364/OE.21.009069}{\detokenize{10.1364/OE.21.009069}}}.

\bibitem[{Pfuhl} \em{et~al.}(2014){Pfuhl} et~al.]{Pfuhl2014}
{Pfuhl}, O.; {Haug}, M.; {Eisenhauer}, F.; {Kellner}, S.;
        {Haussmann}, F.; {Perrin}, G.; {Gillessen}, S.;
        {Straubmeier}, C.; {Ott}, T.; {Rousselet-Perraut}, K.; et al.
\newblock {The fiber coupler and beam stabilization system of the GRAVITY
  interferometer}.
\newblock  \emph{Proc.~SPIE}   \textbf{2014}, \emph{9146},  914623,
\newblock
  doi:{\changeurlcolor{black}\href{https://doi.org/10.1117/12.2055080}{\detokenize{10.1117/12.2055080}}}.
\bibitem[{Perraut} \em{et~al.}(2018){Perraut} et~al.]{Perraut2018}
{Perraut}, K.; Jocou, L.; Berger, J.P.; Chabli, A.; Cardin, V.; Chamiot-Maitral, G.; Delboulbé, A.; Eisenhauer,~F.; Gambérini, Y.; Gillessen, S.; et al.
\newblock Single-mode waveguides for GRAVITY. I. The~cryogenic 4-telescope
  integrated optics beam combiner.
\newblock {\em Astron. Astrophys.} {\bf 2018}, {\em 614},~A70.
\newblock
  doi:{\changeurlcolor{black}\href{https://doi.org/10.1051/0004-6361/201732544}{\detokenize{10.1051/0004-6361/201732544}}}.

\bibitem[{Cl{\'e}net} \em{et~al.}(2010){Cl{\'e}net} et~al.]{Clenet2010}
{Cl{\'e}net}, Y.; Rousset, G.; Eisenhauer, F.; Gillessen, S.; Perrin, G.; Amorim, A.; Brandner, W.; Perraut, K.; Straubmeier, C.
\newblock Dimensioning the Gravity adaptive optics wavefront sensor.
\newblock  \emph{Proc.~SPIE}   \textbf{2010},   \emph{7736},  7364A.
\newblock
  doi:{\changeurlcolor{black}\href{https://doi.org/10.1117/12.856661}{\detokenize{10.1117/12.856661}}}.

\bibitem[{Scheithauer} \em{et~al.}(2016){Scheithauer} et~al.]{Scheithauer2016}
{Scheithauer}, S.; Brandner, W.; Deen, C.; Adler, T.; Bonnet, H.; Bourget, P.; Chemla, F.; Clenet, Y.; Delplancke, F.; Ebert, M.; et al.
\newblock CIAO: wavefront sensors for GRAVITY.
\newblock  \emph{Proc.~SPIE}  \textbf{2016},   \emph{9909},   99092L.
\newblock
  doi:{\changeurlcolor{black}\href{https://doi.org/10.1117/12.2232997}{\detokenize{10.1117/12.2232997}}}.

\bibitem[{Deen} \em{et~al.}(2016){Deen} et~al.]{Deen2016}
{Deen}, C.; Kolb, J.; Oberti, S.; Bonnet, H.; Müller, E.; Hubert, Z.; Zins, G.; Delplancke, F.; Haguenauer,~P.; Pettazzi, L.; et al.
\newblock {System tests and on-sky commissioning of the GRAVITY-CIAO wavefront
  sensors}.
\newblock  \emph{Proc.~SPIE}   \textbf{2016},   \emph{9909},  99092M.
\newblock
  doi:{\changeurlcolor{black}\href{https://doi.org/10.1117/12.2233005}{\detokenize{10.1117/12.2233005}}}.

\bibitem[{Kendrew} \em{et~al.}(2012){Kendrew} et~al.]{Kendrew2012}
{Kendrew}, S.; {Hippler}, S.; {Brandner}, W.; {Cl{\'e}net}, Y.;  {Deen}, C.; {Gendron}, E.; {Huber}, A.; {Klein}, R.; 
	{Laun}, W.; {Lenzen}, R.; et al.
\newblock GRAVITY Coud{\'e} Infrared Adaptive Optics (CIAO) system for the VLT
  Interferometer.
\newblock  \emph{Proc.~SPIE}   \textbf{2012},  \emph{8446},  84467W,
\newblock
  doi:{\changeurlcolor{black}\href{https://doi.org/10.1117/12.926558}{\detokenize{10.1117/12.926558}}}.

\bibitem[{Hippler} \em{et~al.}(2008){Hippler} et~al.]{Hippler2008}
{Hippler}, S.; {Brandner}, W.; {Cl{\'e}net}, Y.; {Hormuth}, F.;
	{Gendron}, E.; {Henning}, T.; {Klein}, R.; {Lenzen}, R.;
	{Meschke}, D.; {Naranjo}, V.; et al.
\newblock Near-infrared wavefront sensing for the VLT interferometer.
\newblock  \emph{Proc.~SPIE}   \textbf{2008},   \emph{7015},~701555,
\newblock
  doi:{\changeurlcolor{black}\href{https://doi.org/10.1117/12.789053}{\detokenize{10.1117/12.789053}}}.

\bibitem[Nijenhuis and Giesen(2005)]{Nijenhuis2005}
Nijenhuis, J.R.; Giesen, P.T.M.
\newblock {A major step forward back in time with the ESO Star Separator
  system}.
\newblock  \emph{Proc.~SPIE}  \textbf{2005},  \emph{5877}, 24--30.
\newblock
  doi:{\changeurlcolor{black}\href{https://doi.org/10.1117/12.618258}{\detokenize{10.1117/12.618258}}}.

\bibitem[{Nijenhuis} \em{et~al.}(2008){Nijenhuis} et~al.]{Nijenhuis2008}
{Nijenhuis}, J.; Visser, H.; de Man, H.; Dekker, B.; Mekking, J.; Kamphues, F.
\newblock {Simultaneous observation of two stars using the PRIMA Star
  Separator}.
\newblock  \emph{Proc.~SPIE}   \textbf{2008},  \emph{7013},  70133F.
\newblock
  doi:{\changeurlcolor{black}\href{https://doi.org/10.1117/12.789943}{\detokenize{10.1117/12.789943}}}.

\bibitem[{Dai} \em{et~al.}(2017){Dai}, {Hippler}, and {Gendron}]{Dai2017}
{Dai}, X.; {Hippler}, S.; {Gendron}, E.
\newblock {Experiments of two pupil lateral motion tracking algorithms using a
  Shack-Hartmann sensor}.
\newblock {\em J.  Mod. Opt.} {\bf 2017}, {\em 64},~127--137.
\newblock
  doi:{\changeurlcolor{black}\href{https://doi.org/10.1080/09500340.2016.1212415}{\detokenize{10.1080/09500340.2016.1212415}}}.

\bibitem[{Finger} \em{et~al.}(2014){Finger} et~al.]{Finger2014}
{Finger}, G.;  Baker, I.; Alvarez, D.; Ives, D.; Mehrgan, L.; Meyer, M.; Stegmeier, J.; Weller, H.J.
\newblock {SAPHIRA detector for infrared wavefront sensing}.
\newblock  \emph{Proc.~SPIE}  \textbf{2014},   \emph{9148},   914817.
\newblock
  doi:{\changeurlcolor{black}\href{https://doi.org/10.1117/12.2057078}{\detokenize{10.1117/12.2057078}}}.

\bibitem[{Finger} \em{et~al.}(2016){Finger} et~al.]{Finger2016}
{Finger}, G.; Baker, I.; Alvarez, D.; Dupuy, C.; Ives, D.; Meyer, M.; Mehrgan, L.; Stegmeier, J.; Weller, H.J.
\newblock {Sub-electron read noise and millisecond full-frame readout with the
  near infrared eAPD array SAPHIRA}.
\newblock  \emph{Proc.~SPIE}   \textbf{2016},   \emph{9909},   990912.
\newblock
  doi:{\changeurlcolor{black}\href{https://doi.org/10.1117/12.2233079}{\detokenize{10.1117/12.2233079}}}.

\bibitem[{Mehrgan} \em{et~al.}(2016){Mehrgan}, {Finger}, {Eisenhauer}, and
  {Panduro}]{Mehrgan2016}
{Mehrgan}, L.H.; {Finger}, G.; {Eisenhauer}, F.; {Panduro}, J.
\newblock {GRAVITY detector systems}.
\newblock  \emph{Proc.~SPIE}  \textbf{2016},   \emph{9907},~99072F.
\newblock
  doi:{\changeurlcolor{black}\href{https://doi.org/10.1117/12.2234731}{\detokenize{10.1117/12.2234731}}}.

\bibitem[{Atkinson} \em{et~al.}(2016){Atkinson} et~al.]{Atkinson2016}
{Atkinson}, D.E.; Hall, D.N.; Baker, I.M.; Goebel, S.B.; Jacobson, S.M.; Lockhart, C.; Warmbier, E.A.
\newblock {Next-generation performance of SAPHIRA HgCdTe APDs}.
\newblock  \emph{Proc.~SPIE}   \textbf{2016}, \emph{ 9915},  99150N.
\newblock
  doi:{\changeurlcolor{black}\href{https://doi.org/10.1117/12.2234314}{\detokenize{10.1117/12.2234314}}}.

\bibitem[{Finger} \em{et~al.}(2019){Finger} et~al.]{Finger2019}
{Finger}, G.; Baker, I.; Alvarez, D.; Eisenhauer, F.; Hechenblaikner, G.; Ives, D.; Mehrgan, L.; Meyer, M.; Stegmeier, J.; Weller, H.J.
\newblock On-sky performance verification of near infrared eAPD technology for
  wavefront sensing at ground based telescopes, demonstration of e-APD pixel
  performance to improve the sensitivity of large science focal planes and
  possibility to use this technology in space.
\newblock  \emph{Proc.~SPIE}   \textbf{2019}, \emph{11180},  111806L.
\newblock
  doi:{\changeurlcolor{black}\href{https://doi.org/10.1117/12.2536156}{\detokenize{10.1117/12.2536156}}}.

\bibitem[{Goebel} \em{et~al.}(2018){Goebel} et~al.]{Goebel2018}
{Goebel}, S.B.; {Jacobson}, S.M.; {Lockhart}, C.;
         {Warmbier}, E.A.
\newblock {Overview of the SAPHIRA detector for adaptive optics applications}.
\newblock {\em J. Astron. Telesc. Instrum. Syst.}
  {\bf 2018}, {\em 4},~026001,
  doi:{\changeurlcolor{black}\href{https://doi.org/10.1117/1.JATIS.4.2.026001}{\detokenize{10.1117/1.JATIS.4.2.026001}}}.

\bibitem[{Fowler} and {Gatley}(1990)]{Fowler1990}
{Fowler}, A.M.; {Gatley}, I.
\newblock {Demonstration of an Algorithm for Read-Noise Reduction in Infrared
  Arrays}.
\newblock {\em  Astrophys. J.} {\bf 1990}, {\em 353},~L33.
\newblock
  doi:{\changeurlcolor{black}\href{https://doi.org/10.1086/185701}{\detokenize{10.1086/185701}}}.

\bibitem[{Baker} \em{et~al.}(2016){Baker}, {Maxey}, {Hipwood}, and
  {Barnes}]{Leonardo2016}
{Baker}, I.; {Maxey}, C.; {Hipwood}, L.; {Barnes}, K.
\newblock {Leonardo (formerly Selex ES) infrared sensors for astronomy: present
  and future}.
\newblock  \emph{Proc.~SPIE}  \textbf{2016},  \emph{9915},  991505.
\newblock
  doi:{\changeurlcolor{black}\href{https://doi.org/10.1117/12.2231079}{\detokenize{10.1117/12.2231079}}}.

\bibitem[{Lacour} \em{et~al.}(2019){Lacour} et~al.]{Lacour2019}
{Lacour}, S.; {Dembet}, R.; {Abuter}, R.; {F{\'e}dou}, P.;
         {Perrin}, G.; {Choquet}, {\'E}.; {Pfuhl}, O.;
         {Eisenhauer}, F.; {Woillez}, J.; {Cassaing}, F.; et al.
\newblock {The GRAVITY fringe tracker}.
\newblock {\em Astron. Astrophys.} {\bf 2019}, {\em 624},~A99,
\newblock
  doi:{\changeurlcolor{black}\href{https://doi.org/10.1051/0004-6361/201834981}{\detokenize{10.1051/0004-6361/201834981}}}.

\bibitem[{White} \em{et~al.}(1974){White}, {Lampe}, {Blaha}, and
  {Mack}]{White1974}
{White}, M.H.; {Lampe}, D.R.; {Blaha}, F.C.; {Mack}, I.A.
\newblock {Characterization of surface channel CCD image arrays at low light
  levels}.
\newblock {\em IEEE J. Solid-State Circuits} {\bf 1974}, {\em
  9},~1--12.
\newblock
  doi:{\changeurlcolor{black}\href{https://doi.org/10.1109/JSSC.1974.1050448}{\detokenize{10.1109/JSSC.1974.1050448}}}.

\bibitem[{Alessandri} \em{et~al.}(2016){Alessandri} et~al.]{Alessandri2016}
{Alessandri}, C.;{Abusleme}, A.; {Guzman}, D.; {Passalacqua}, I.;  {Alvarez-Fontecilla}, E.; {Guarini}, M.
\newblock Optimal CCD readout by digital correlated double sampling.
\newblock {\em Mon. Not. R. Astron.~Soc.} {\bf 2016},
  {\em 455},~1443--1450,
\newblock
  doi:{\changeurlcolor{black}\href{https://doi.org/10.1093/mnras/stv2410}{\detokenize{10.1093/mnras/stv2410}}}.

\bibitem[{Cruz de la Torre} and {de Vicente Albendea}(2018)]{CruzDeLaTorre2018}
{Cruz de la Torre}, C.; {de Vicente Albendea}, J.
\newblock {Digital correlated double sampling CCD readout characterization}.
\newblock \emph{ Proc.~SPIE}   \textbf{2018},  \emph{10709},  1070920.
\newblock
  doi:{\changeurlcolor{black}\href{https://doi.org/10.1117/12.2312498}{\detokenize{10.1117/12.2312498}}}.
\bibitem[{Fusco} \em{et~al.}(2004){Fusco} et~al.]{Fusco2004}
{Fusco}, T.; Ageorges, N.; Rousset, G.; Rabaud, D.; Gendron, E.; Mouillet, D.; Lacombe, F.; Zins, G.; Charton,~J.; Lidman, C.; et al.
\newblock NAOS performance characterization and turbulence parameters
  estimation using closed-loop data.
\newblock  \emph{Proc.~SPIE}  \textbf{2004},   \emph{5490},  118--129.
\newblock
  doi:{\changeurlcolor{black}\href{https://doi.org/10.1117/12.563009}{\detokenize{10.1117/12.563009}}}.

\bibitem[{Snyder} \em{et~al.}(2016){Snyder}, {Srinath}, {Macintosh}, and
  {Roodman}]{Snyder2016}
{Snyder}, A.; {Srinath}, S.; {Macintosh}, B.; {Roodman}, A.
\newblock {Temporal characterization of Zernike decomposition of atmospheric
  turbulence}.
\newblock  \emph{Proc.~SPIE}   \textbf{2016},  \emph{ 9906},  990642.
\newblock
  doi:{\changeurlcolor{black}\href{https://doi.org/10.1117/12.2234362}{\detokenize{10.1117/12.2234362}}}.

\end{thebibliography}
%

\end{document}